\documentclass[runningheads]{llncs}

\usepackage[CJKbookmarks=true,bookmarksopen=true,colorlinks,citecolor=black,linkcolor=black,anchorcolor=black,urlcolor=black] {hyperref}

\usepackage{url}
\usepackage{multirow}

\usepackage[english]{babel}
\usepackage{hyphenat}

\usepackage{amsmath}

\usepackage{graphicx}
\graphicspath{{./}}
\usepackage{subfig}
\begin{document}
\title{IoTDataBench: Extending TPCx-IoT for Compression and Scalability}
\author{Yuqing Zhu\inst{1,2}\orcidID{0000-0001-5431-8581} \and
Yanzhe An\inst{2,3} \and
Yuan Zi\inst{2,3} \and
Yu Feng\inst{2,3} \and
Jianmin Wang\inst{1,2,3}}
\authorrunning{Y. Zhu et al.}
\institute{BNRIST, Tsinghua University, Beijing 100084, China \\
\email{\{zhuyuqing, jimwang\}@tsinghua.edu.cn}\and
NELBDS, Tsinghua University, Beijing 100084, China \\
\email{\{ayz19, ziy20, y-feng19\}@mails.tsinghua.edu.cn}\and
School of Software, Tsinghua University, Beijing 100084, China
}
\maketitle              % typeset the header of the contribution
\begin{abstract}
%We present results and lessons learned in practicing TPCx-IoT benchmarking for a real-world scenario. We find that more system characteristics need to be benchmarked for its application to real-world use cases. We introduce an extension to the TPCx-IoT benchmark, covering fundamental requirements of time-series data management for IoT infrastructure. We characterize them as data compression and system scalability. To evaluate these two important features of IoT databases, we develop IoTDataBench and update four aspects of TPCx-IoT, i.e., data generation, workloads, metrics and test procedures. Preliminary experimental results show systems that fail to effectively compress data and flexibly scale negatively affect the redesigned metrics, while systems with high compression ratios and system scalability are rewarded in the final metrics. Such systems have the ability to scale up computing resources on demand and can thus save dollar costs.

We present a record-breaking result and lessons learned in practicing TPCx-IoT benchmarking for a real-world use case. We find that more system characteristics need to be benchmarked for its application to real-world use cases. We introduce an extension to the TPCx-IoT benchmark, covering fundamental requirements of time-series data management for IoT infrastructure. We characterize them as data compression and system scalability. To evaluate these two important features of IoT databases, we propose IoTDataBench and update four aspects of TPCx-IoT, i.e., data generation, workloads, metrics and test procedures. Preliminary evaluation results show systems that fail to effectively compress data or flexibly scale can negatively affect the redesigned metrics, while systems with high compression ratios and linear scalability are rewarded in the final metrics. Such systems have the ability to scale up computing resources on demand and can thus save dollar costs.

\end{abstract}
%
% InfluxDB, Apache HBase and IoTDB
%
\section{Introduction}

The Internet of Things (IoT) are increasingly attracting attentions from both industry and academia. The increasing number of IoT devices and sensors has driven IoT data management system, i.e., time-series database, to be the most popular among all types of databases~\cite{tsdbengines}. Hence, there are increasing demands to benchmark and compare various design and implementations of time-series databases.

TPCx-IoT~\cite{tpcxiotAnalysis} is an industrial standard for benchmarking IoT data management systems. It targets at the databases deployed for the Internet-of-Things architecture. It addresses the important aspects in IoT data management such as intensive data ingestion and time-range based queries. In comparison to alternative benchmarks for IoT data management like TSBS~\cite{tsbs}, TPCx-IoT can increase the generated workload by horizontal scale-out. This is important for IoT data management, as IoT data workload features intensive writes that are not common in other workloads such as key-value or relational data. The intensive write demands can be as high as many tens of millions of points per second.

Despite the benefits of TPCx-IoT, we find that it has not addressed some key aspects of IoT data management scenarios. On running TPCx-IoT benchmark over our open-sourced distributed IoT database, IginX\footnote{https://github.com/thulab/IginX} over Apache IoTDB\footnote{https://github.com/apache/IoTDB}, we achieved a record-breaking result, \textbf{4,116,821 IoTps}. When communicating the result to our users for a use case, we find that the result is not as useful as expected, because data compression and system scalability are not tested.

Data compression is very important for IoT data management, as the data volume is unprecedentedly huge. The target deployment environment of IoT database might not install computing resources to hold such a large volume of data in its original size. Data compression is the fundamental measure to reduce storage requirement. Moreover, the limited computing resource of the target deployment environment requires processing before synchronization with the cloud environment. Such processing does not include the time-range based queries, but also aggregation queries~\cite{tsbs,smartbench}, which are extremely common for detailed data from IoTs. These aggregation-based workloads are considered in some common tests for IoT data management~\cite{tsbs,smartbench}, but not in TPCx-IoT.

Furthermore, IoT devices can increase as time passes, leading to so-called cardinality problem that is difficult to handle~\cite{tscardinality}. Due to the inherently limited computing resources of a single server, system scalability is the inevitable measure to handle this requirement of applications. It is in fact a pressing requirement when considering the highly increasing velocity and volume of IoT data, due to the potentially fast increase of IoT devices and sensors.

Existent benchmarks commonly used~\cite{tsbs} have not covered or even considered data compression and database system scalability. Neither have academic benchmarks~\cite{smartbench}. While each of these benchmarks has its own advantage respectively, we amalgamate their features and add two new features, i.e., data compression and system scalability tests. As TPCx-IoT is currently the industrial standard for IoT database benchmarking, we extend TPCx-IoT and propose the new benchmark for IoT data management, IoTDataBench.

IoTDataBench features data compression test and database scalability test. Data compression is largely related to data type and data distribution. To benchmark data compression capability properly, IoTDataBench integrates a data modeling framework such that various real-world time-series data can be plugged into the framework to train models for later data generation. We extend the TPCx-IoT benchmark procedure to include the new data generation component. We also adapt the workload execution run to include the system scalability test procedure. Finally, we update the benchmark metrics to incorporate the results of data compression test and system scalability test. In sum, we make the following contributions:
\begin{itemize}
  \item We present a record-breaking result and the lessons learned in practicing TPCx-IoT benchmarking for a real-world use case.
  \item We make an apple-to-apple comparison between common benchmarks for IoT and time-series data management, summarizing their features and shortcomings.
  \item We propose IoTDataBench by extending the TPCx-IoT benchmark to include data compression test and system scalability test.
  \item We implement IoTDataBench based on TPCx-IoT, updating metric computation, data generation, workload generation and test procedure.
%  \item We present a discussion regarding the data compression and the scalability features of popular time-series databases that are commonly used for IoT data management.
  \item Our preliminary evaluation results validate the design of IoTDataBench and indicate the future directions for improving IoT database systems.
\end{itemize}
\begin{figure}[t]
	\centering%\captionsetup{width=.5\textwidth}%
	\includegraphics[width=\textwidth]{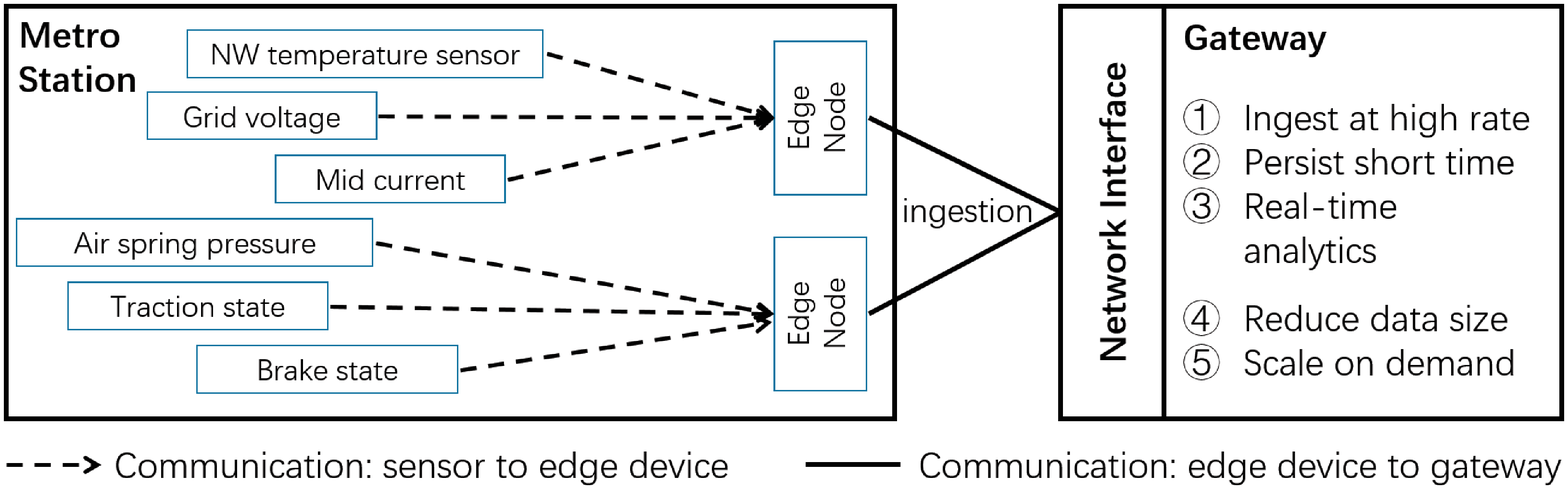}%
	\caption{Overview of a gateway architecture serving metro trains.}\vspace{-12pt}%
	\label{fig:gateway} %
\end{figure}
\section{Use Case: Benchmarking for Train Monitoring}

In this section, we present a use case when we try to exercise TPCx-IoT for selecting the database for an IoT data management scenario. We have analyzed the benchmarking requirements and studied the in-depth analysis of the TPCx-IoT benchmark~\cite{tpcxiotAnalysis}. In the following, we present our results in practicing TPCx-IoT. In the process, we learned a few lessons that make us to extend TPCx-IoT and develop IoTDataBench.

One of our user, Shanghai Metro, has a monitoring system for its subway trains. The monitoring system must support dashboard-like functionality. It must keep all the monitoring data received from all sensors on all trains by MQTT protocol. As shown in Figure~\ref{fig:gateway}, it is facing a scenario resembling the gateways in a typical IoT infrastructure~\cite{tpcxiotAnalysis}. As a result, we try benchmarking our database by TPCx-IoT, an industrial benchmark for IoT databases. We aim to see how well our database can go and whether it satisfies the user's requirements.

\subsection{Benchmarking Result and Settings}

\begin{figure}[t]
	\centering%\captionsetup{width=.5\textwidth}%
	\includegraphics[width=0.9\textwidth]{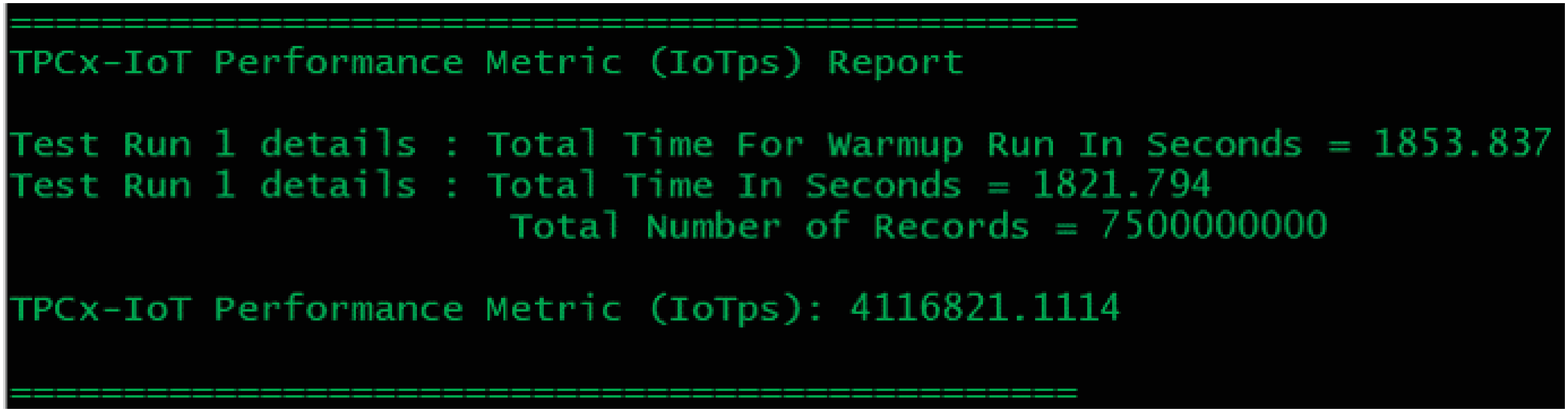}%
	\caption{The benchmarking result.}\vspace{-12pt}%
	\label{fig:result} %
\end{figure}

In our benchmarking test, the final measured run lasts for 1821.794 seconds, leading to the record-breaking result of \textbf{4,116,821 IoTps}, as shown in Figure~\ref{fig:result}.

We run TPCx-IoT benchmark tests on IginX over Apache IoTDB. While IoTDB is a standalone database for IoT time-series data management, IginX is the middleware that can manage multiple instances of standalone IoT time-series databases to form an elastic cluster. As long as a standalone IoT time-series database implements the database interface, it can be integrated into and managed by IginX. Currently, IginX supports IoTDB and InfluxDB. Both IginX and IoTDB are open-sourced. The tests are run on a cluster of five servers, each with the following configuration:
\begin{itemize}
  \item 4 Intel(R) Xeon(R) CPU Platinum 8260 clocked at 2.40GHz, each with 24 cores and 48 threads.
  \item 768GB RAM
  \item Mellanox HDR (200 Gbps per server node)
  \item 1.6TB Nvme Disk
\end{itemize}

We used the following tuning parameters for IginX and IoTDB:
\begin{itemize}
  \item IginX, sessionPoolSize=130
  \item IginX, asyncExecuteThreadPool=100
  \item IginX, syncExecuteThreadPool=120
  \item IoTDB, mlog\_buffer\_size=16777216
  \item IoTDB, enable\_mem\_control=false
  \item IoTDB, max\_deduplicated\_path\_num=10000
  \item IoTDB, compaction\_strategy=NO\_COMPACTION
  \item IoTDB, enable\_unseq\_compaction=false
  \item IoTDB, enable\_last\_cache=false
\end{itemize}

For TPCx-IoT, we set the number of clients to be five, each of which co-locates with one IginX instance and one IoTDB instance on a server node. Each client is set with 12 threads. The database records count is 7.5 billion. Except for environmental-setting related parameters, no modifications are made to the original TPCx-IoT distribution files.
\subsection{Learned Lessons}

While our benchmarking results are promising, our users find that the results are not as helpful as expected for their application scenarios. In fact, after fully analyzing the requirements in the use case and the TPCx-IoT specification, we find that TPCx-IoT falls short in the following aspects.

First, TPCx-IoT is not considering data compression. Sensors on the trains are generating more than 10TB data every month. While the company would like to preserve all the data for future analysis, the storage cost is causing a burden for this non-Internet company. As time-series database is suited for managing IoT data and performs well in data compression, the company would like to compare which database performs best in data compression and thus reduces storage costs most. Besides, data compression is also important to reduce bandwidth consumption when transferring data from gateways back to the cloud systems.

Second, TPCx-IoT does not take the database scalability into account, although it has well considered the scalability of workload generation so as to simulate the fast-ingestion workload pattern of IoT scenarios. Shanghai Metro has to scale out the whole database system every month for more storage space, since sensors are generating more than 10TB data per month~\cite{dhtlb}. Unfortunately, the company finds that the database system in use takes longer and longer time for the scale-out process, during which the database performance degrades severely. Besides, when new lines are opened or new sensors are added to trains, the database system will also need to scale out accordingly. But TPCx-IoT lacks scalability tests on database system.

Third, TPCx-IoT has not used real-world data in the benchmarking tests. The distribution of data in the workload is key to measuring the performance of IoT databases. On the one hand, data compression is highly correlated with data distribution. For example, sensor data without white noise can be compressed much better than one with. Besides, data types also matter, because numerics can be encoded but strings normally cannot. On the other hand, while TPCx-IoT uses values of exactly 1KB and considers only evenly-spacing time-series data, real-world data are commonly sent back in the original size without padding to the same size, and unevenly-spacing data are also common in IoT scenarios. These factors have also impacts on the results of data compression. Furthermore, while some time-series databases have devised specific accelerating mechanisms for data queries, data type and distribution also play an important role in affecting the performance of queries.

From the above analyses about the use case, we learn that 1) the system's capability in scaling-out must be measured to meet the increased workloads that are highly possible in future; 2) users need to be informed about the system performance on data compression to estimate their future cost of their storage space; and, 3) real-world data are needed for a valid evaluation result of the IoT data management system.

The use case of Shanghai Metro is among the common scenarios for IoT data management, e.g., aeroplane flight test monitoring and truck monitoring. As a result, we believe that TPCx-IoT need be extended to include data compression and system scalability. Besides, data types and distribution must be taken into account instead of using padding bytes.

\section{Related Works}

As for benchmarking IoT databases, three representative benchmarks exist for common usage. The first is TPCx-IoT~\cite{tpcxiotAnalysis}, which is proposed by the TPC organization as a standard. The second is TSBS, which is first setup by the InfluxDB project and then improved by the TimescaleDB project. While InfluxDB is ranked as the most popular project and TimescaleDB the tenth popular, TSBS has become a defacto standard benchmark for open-source time-series database. As IoT data are typically time-series data, TSBS~\cite{tsbs} can also be used for IoT database benchmarking. The third is SmartBench, which is an academic work. SmartBench~\cite{smartbench} mainly considers comprehensive analytical requirements of IoT data. Therefore, it has a more inclusive query workloads.

TPCx-IoT mainly addresses three key aspects of IoT data management. The first is whether intensive writes can be generated. Intensive writes are typical of IoT data management scenarios. This features differ greatly from KV or relational data management. To settle this, TPCx-IoT designs a highly scalable architecture that can distribute the workload generator across multiple nodes. The second is whether typical queries can be simulated. Time-range queries are common in IoT data processing. TPCx-IoT mainly simulate such query workloads. The third is whether data reliability is guaranteed. TPCx-IoT checks the number of replicas in the benchmarking procedure. Besides, it also sends writes and queries on the same data to different clients for verification.

We compare the three benchmarks in Table~\ref{tbl:comparison}, along with our proposal of IoT data benchmark, i.e., IoTDataBench. As for query workloads, while TPCx-IoT is expected to cover aggregation queries, we find its official implementation only supports time-range queries. But TSBS and SmartBench cannot support mixed read-write workloads. In comparison, TPCx-IoT adopts the YCSB framework and can flexibly generate any mixture of read-write workloads. As for scalability, TSBS and SmartBench lacks both benchmark scalability and tests on database scalability. TPCx-IoT has great benchmark scalability, but cannot test the scalability of database. None of the three benchmarks have considered benchmarking the database' capability on data compression. Extending from TPCx-IoT, IoTDataBench inherits its features and advantages. Besides,  IoTDataBench has all the desired features of scalability test on database and data compression test.
\begin{table}[t]
\caption{Comparing benchmarks for IoT data management.}%
\centering
\scriptsize
\label{tbl:comparison}
\begin{tabular}{|l|l|c|c|c|c|}
\hline
\multirow{2}{*}{\bf Benchmarks} & \multicolumn{1}{|c|}{\multirow{2}{*}{\bf Queries}} & \multirow{2}{*}{\bf Mixed R/Ws} & \multicolumn{2}{|c|}{\bf Scalability}&\multirow{2}{*}{\bf Compression}\\
\cline{4-5}
& & & Benchmark & Database& \\
\hline
TSBS& Time-range and aggregation & NO& NO & NO& NO\\
SmartBench & Time-range and aggregation & NO& NO & NO& NO\\
TPCx-IoT& Time-range-only & YES& YES & NO& NO\\
IoTDataBench& Time-range and aggregation & YES& YES & YES & YES\\
\hline
\end{tabular}\vspace{-12pt}
\end{table}
\section{IoTDataBench: A TPCx-IoT Evolution}

IoTDataBench extends the use cases supported by TPCx-IoT to more general scenarios, where new IoT devices can be added in an ad-hoc way. In such scenarios, workloads will increase accordingly, requiring the database to have the capability to scale-out in an on-demand manner. The use case that motivates our design of IoTDataBench also represents the use case when system scalability is demanded for storage expansion.
\begin{figure}[t]
	\centering%\captionsetup{width=.5\textwidth}%
	\includegraphics[width=0.75\textwidth]{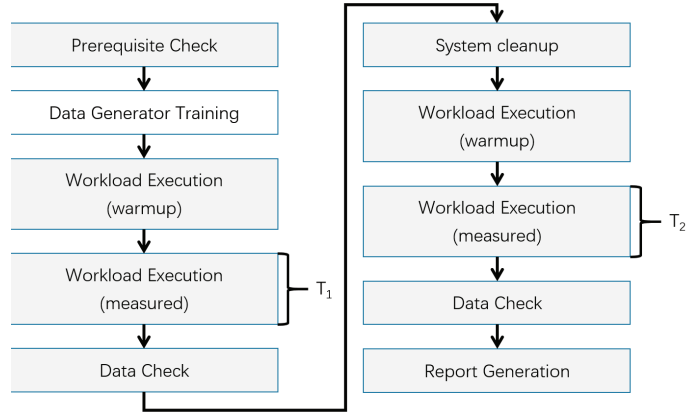}%
	\caption{IoTDataBench benchmarking procedure.}\vspace{-12pt}%
	\label{fig:procedure} %
\end{figure}
\subsection{Benchmarking Procedure}

IoTDataBench has a benchmarking procedure resembling TPCx-IoT's execution rules. IoTDataBench modifies the execution rules to include a data generator modeling phase such that real-world data can be incorporated in the benchmarking tests. Although execution rules might require long lasting discussions to reach an agreement, the benchmarking procedure of IoTDataBench updates the execution rules only for real-world data inclusion. This data generator modeling phase is not in the measured steps. Hence, the addition of this phase does not change the main thread of TPCx-IoT's execution rules.

As illustrated in Figure~\ref{fig:procedure}, a typical benchmark run of IoTDataBench consists of two benchmark iterations. In each benchmark iteration, the warmup phase of workload execution runs stable workloads with mixed ingestion (write) and query (read) as in TPCx-IoT; and, the measured run of workload execution differs from the warmup phase by including database scale-out test procedures. Besides, data check is carried out in each iteration for data validation. The first iteration has an extra phase for data generator modeling, which trains a model for data generation used in both iterations. After the first iteration, the system is cleaned up for the second iteration. Before the benchmark run, a set of prerequisite checks are carried out for specification-conformity tests. After the benchmark run, results are summarized with a report generated.

During benchmark tests, IoTDataBench records the total volume $S_i$ of ingested data and collects the final on-disk storage size $S_d$ of the database system under test. These metrics are collected for data compression test. Besides, IoTDataBench records the number of ingested data points for either subphase of a workload execution run, i.e., $n_0$ for the stable phase and $n_s$ for the scale-out phase. The total number $N_p$ of ingested data points is the sum of $n_0$ and $n_s$.

As in TPCx-IoT, IoTDataBench accepts the number of simulated sensors $m_s$ as the input. We can compute the average number of ingested points per second by $\frac{N_p}{m_s}$. Following TPCx-IoT's rules, the minimal average number of ingested points per second is required to be above $20$. While the runtimes of both workload execution runs are collected, represented as $T_1$ and $T_2$, each run of workload execution must exceed $1800s$.

\subsection{Data Model and Data Generation}

TPCx-IoT generates random bytes for writes and reads, while TSBS pre-generates random numerics for benchmarking. In practice, numerics dominate the data types of time-series values. For DevOps scenarios, strings are also common for logging purposes. IoTDataBench considers both literals and numerics.

IoTDataBench provides three methods for data generation. The first is generating numerics by stochastic distributions. Currently, TPCx-IoT supports constant, uniform, Zipfian, and histogram random number generation. These generators are mainly for integers. Besides, they do not follow the real-world distributions of IoT data. Therefore, we provide data generator for floating point numbers. We integrate multiple generators for common distributions, e.g., Poisson, Pareto, and exponential distributions. These generators can be used for both timestamps and values.

The second method is to let users input a sample of their target dataset and to generate more data by copying the sample periodically. Data generated in this way naturally follow the distribution of real-world data. Besides, the common periodicity of time-series data is instantiated in the generated data.

The third method also allows users to input a sample of their target dataset. Instead of just copying the original data, this method trains a data generation model by the data sample. Then the model is used for data generation. While this method can take longer time for data generation than other methods, the time taken by data generation phase is not counted in the running time of benchmark tests.

Currently, IoTDataBench exploits Poisson, Pareto and exponential distributions to generate floating point numbers for tests~\cite{summarystore}. However, the other two methods are also provided with configurable parameters for users such that it can be applied to broader range of applications. Furthermore, to enable the generation of periodical patterns, IoTDataBench keeps 10 sets of 6710886 points, leading to approximately 1GB data. Each set is recursively passed to some client threads as input to form periodical patterns.

\subsection{Workload Generation: Ingestion and Query}

\textbf{Ingestion workloads.} We integrate the workload generator with the ingestion procedure. Different from TPCx-IoT having exactly 1KB values for each write, IoTDataBench generates variable-type values by the trained data generator. Such values includes integers, floating numbers, and strings. The different data types of values are to examine the data compression capability of system under tests. Besides evenly-spaced time-series data, IoTDataBench also considers unevenly-spaced time-series data with stochastically distributed timestamps.

\textbf{Query workloads.} We integrate IoTDataBench with query types considered in counterparts, i.e., TSBS and SmartBench. These queries are specific to IoT data. Besides, data compression can also influence the query performance. The system scalability test is also possible to influence the overall system performance. Queries are generated from the following four query templates, which represent typical dash-board-like queries~\cite{smartbench,tsbs}:
\begin{enumerate}
  \item Time-range based observations $(S\subseteq Sensors, t_s, t_e)$: selects observations from sensors in the list of sensors $S$ in a time range of $[t_s,t_e]$.
  \item Aggregation based statistics $(S\subseteq Sensors, F, t_s, t_e)$: retrieves statistics (e.g., average, max, min, first, last), based on functions specified in $F$ during the time range $[t_s,t_e]$.
  \item Down-sampled observations $(S\subseteq Sensors, t_u, t_s, t_e)$: selects downsampled observations by the time unit $t_u$ from sensors in the list of sensors $S$ in a time range of $[t_s,t_e]$.
  \item Filtered observations $(S\subseteq Sensors, cond, t_s, t_e)$: selects observations that satisfied the condition $cond$ from sensors in the list of sensors $S$ in a time range of $[t_s,t_e]$. $cond$ is in the form of $<s_v> \theta <value>$, where $s_v$ is the sensor name, $value$ is in its range, and $theta$ is a comparison operator, e.g., equality.
\end{enumerate}

IoTDataBench groups the queries of TPCx-IoT, TSBS and SmartBench to form four representative query types. As compared to TPCx-IoT, the latter two types of query workloads are new. While it is possible to implement the latter three types of queries based on the first type of queries, it is much costly than the implementation of directly supporting these queries. Directly support of the latter three query types are typical of databases designed for IoT data management.

\subsection{Database Scalability Test}

IoTDataBench features database scalability test. This test is run within the workload execution process. Each measured run of workload execution is divided into a stable phase and a scale-out phase. The stable phase lasts for half of the warmup runtime. Then, the system is scaled out to \emph{one} node by the supplied database command. Afterwards, one more TPCx-IoT client is added for generating more workloads. Here, IoTDataBench requires the scale-out process to be node-by-node, because this fits better with real-world application scenarios. In case that a system cannot support the scale-out process, the system can be specified with \texttt{non-scalable}. It is required that each round of workload execution must run for at least 30 minutes.

On generating workloads for database scalability test, there exists the key problem of deciding the number of records for each client. With $k$ clients configured for workload generation, one client is used for scalability test. For the $N_p$ database records, each of the $k$ clients is assigned with $2N_p/(2k-1)$ records, while the remaining client is assigned with $N_p/(2k-1)$ records. As the stable phase lasts for only half of the warmup runtime, the $k-1$ clients should finish generating half of their workloads before the scalability test. Hence, all clients are generating approximately the same workloads, again following the original uniform pattern of client workload generation in TPCx-IoT.

We simulate the scale-out scenario for situations when new IoT devices are installed and connected to the system as planned. We require the system to automatically scale out because automatic scalability of system can save maintenance cost. Besides, research on edge resiliency has attracted attentions. We believe benchmarks for IoT data management should consider aspects for the future. The increase ingestion workload exerts pressures on the system's capability in handling high cardinality, as well as in handling intra-system workloads during scale-out process.
\begin{figure}[t]
	\centering%\captionsetup{width=.5\textwidth}%
	\includegraphics[width=0.9\textwidth]{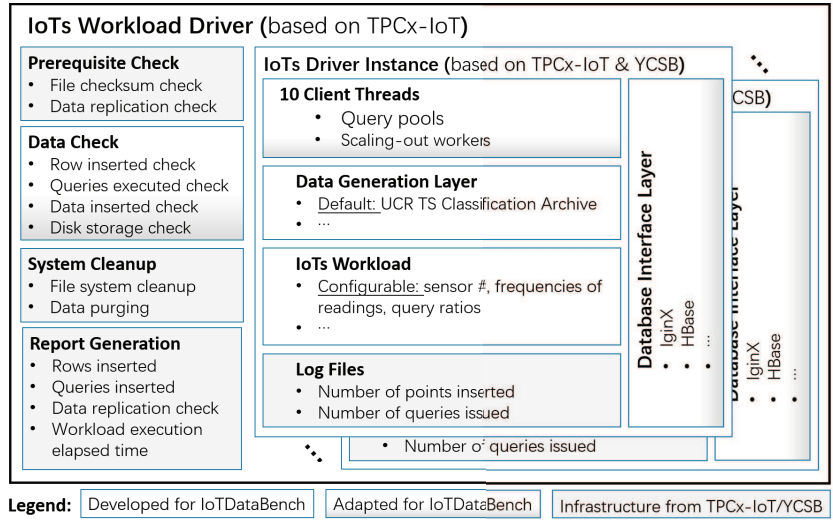}%
	\caption{IoTDataBench Architecture.}\vspace{-12pt}%
	\label{fig:arch} %
\end{figure}
\subsection{Benchmark Driver Architecture}

IoTDataBench inherits the same benchmarking architecture as TPCx-IoT~\cite{tpcxiotAnalysis}, with a few noticeable differences. Figure~\ref{fig:arch} illustrates the overall architecture of IoTDataBench. The workload driver is responsible for running the entire workload. Before initiating the main benchmarking process, some prerequisite checks are carried out. Then, IoTDataBench workload driver also accepts three arguments for initiating its driver instances, i.e., \emph{the number $n_i$ of driver instances} and \emph{the total number $N_p$ of points}. The workload driver spawn $n_i$ driver instances accordingly. After the measured workload execution phase finishes, the workload driver initiates data check, system cleanup and report generation as needed.

In Figure~\ref{fig:arch}, white components are developed for IoTDataBench, and components with gradually changing color are adapted for IoTDataBench. The shaded components are directly inherited from TPCx-IoT and YCSB. IoTDataBench implements two new components. The first is a data generation component for a data modeling phase such that data can be generated in an ad-hoc manner. The second is a new workload generator for IoTDataBench to suit the different application scenarios for IoT data management.

IoTDataBench adapts three components from TPCx-IoT, i.e., data check, thread pool and database interface layer. Two more checks are added to data check phase. They are \emph{data inserted check} for checking the ingested data volume $S_i$ in bytes, and \emph{disk storage check} for checking the total storage size $S_d$ consumed by the database for the test. Results from these two checks are used to compute the data compression metric. IoTDataBench adds the function of scaling out workers to the three management component. This addition is for the scalability test of the benchmark. As for database interface layer, IoTDataBench extends the original database interface to accept more IoT data-specific processing operations such as aggregation queries.

\subsection{Metrics}

IoTDataBench extends TPCx-IoT to measure two more aspects, i.e., data compression ratio and scalability. To follow TPC benchmarks' virtue of publishing only one performance metric and one price/performance metric, we incorporate these two aspects into the original two metrics of TPCx-IoT respectively, i.e., performance metrics (IoTps) and price-performance metrics (\$/IoTps).

\textbf{IoTps:} The performance metric reflects the effective ingestion rate in seconds that the SUT can support during a 30-minute measurement interval, while executing real-time analytic queries and scaling out once. The total number $N_p$ of points ingested into the database is used to calculate the performance metric. For the two measured runs, the performance run is defined as the one with the longer running time $T_i=max(T_1,T_2)$. IoTps is calculated as follows:
\begin{equation}
IoTps = \frac{N_p}{T_i}
\end{equation}

As $N_p$ is summarized from two parts, i.e., $n_0$ and $n_s$, it naturally reflects the system's capability to scale out. As data compression must be exerted in the process of workload execution, the compression performance is also considered in a comprehensive way.

\textbf{\$/IoTps:} The price-performance metric reflects the total cost of ownership per unit IoTps performance. To incorporate the consideration of data compression and scalability, IoTDataBench makes the following adaptation. First, the cost of the storage component is not considered in a whole. Rather, given the cost $C_0$ of the storage component and its storage space $S_0$, the final cost of the storage component is computed as $C_0\frac{S_d}{S_i}$, where $S_0$ must be larger than or equal to $S_i$. Hence, a higher compression ratio, i.e., $r=\frac{S_i}{S_d}$ leads to a lower storage cost.

The unit cost of storage can be computed as $\frac{C_0S_d}{N_pS_i}$. To support $IoTps$ workload, the storage cost should be compute as $IoTps\frac{C_0S_d}{N_pS_i}$. Given that storage cost takes up more and more portion of the total system cost as time passes, we further augment the storage cost by a weight $w_i=2*24*365=17520$. This augmentation weight is given by assuming that the system will keep data for one-year long and that the system will run for 24 hours in the whole year.

Second, we further take into account scalability in the price-performance metric. As the system can be consisted of different number of nodes at different stage of the workload execution run, we compute the system cost by a weighted sum. Assume the system before scale-out costs $c_0$ and that after scale-out costs $c_s$. The total cost $C_S$ of the system is computed as $\frac{1}{2}c_0+\frac{1}{2}c_s$.

\section{Evaluation}

Metric is paramount in the design of a benchmark. We present our preliminary evaluation of the extended metrics, along with a description of our IoTDataBench implementation. Preliminary evaluation results show systems that fail to effectively compress data or flexibly scale can negatively affect the redesigned metrics, while systems with high compression ratios and linear scalability are rewarded in the final metrics. Such systems have the ability to scale up computing resources on demand and can thus save dollar costs.

\subsection{Implementation}

For the evaluation of IoTDataBench, we implemented data generation and workload generation extensions within the YCSB~\cite{ycsb} codes extended by TPCx-IoT. As for data generation, we added two data generation classes for Poisson and Pareto distributions. Following TPCx-IoT's implementation, we modified the source codes of the \texttt{CoreWorkload}, \texttt{DB}, and \texttt{DBWrapper} class. While time-range query was already implemented, we implemented the other three types of queries by adding two methods, with the aggregation and the down-sample queries supported in one method.

The scalability test and the metric computation were implemented by adapting the shell scripts, including \texttt{TPC-IoT-master.sh} and \texttt{TPC-IoT-client.sh}. The disk storage check was implemented as a new shell script to be executed at each of the database server nodes.

\subsection{Performance Metric Evaluation}

We first evaluate whether the performance metric can reflect the scalability feature of the system under test. Let $n_0$ and $n_s$ be the total number of points ingested during the stable phase and the scale-out phase of the workload execution procedure respectively. Besides, let $t_0$ and $t_s$ be the time taken by the two phases respectively. Assume the system under test is scaled from $m$ nodes to $m+1$ nodes. If the system is linearly scalable, according to the new performance definition, we have:\vspace{-6pt}
\begin{eqnarray}
n_s=\frac{n_0}{t_0}\frac{m+1}{m}w_s, w_s=1\\
N_p=n_0+n_s\\
T_i=t_0+t_s\\
IoTps=\frac{1}{t_0+t_s}(t_s\frac{n_0}{t_0}\frac{m+1}{m}w_s+n_0)=\frac{N_p}{T_i}\label{eqn:iotps}
\end{eqnarray}\vspace{-12pt}

According Eq.~\eqref{eqn:iotps}, $w_s$ reflects the scalability property of the system under test, while $m$ rewards systems consisting of powerful nodes in smaller numbers. This complies with the design assumption of TPCx-IoT. Assume the system under test can achieve $4100$ \emph{KIoTps} in TPCx-IoT with different numbers of nodes. We plot the corresponding performances for systems with different numbers of nodes, assuming linear scalability, i.e., $w_s=1$, and non-linear scalability, i.e., $w_s=0.9$. For non-linear systems, the linearity decreases as the system size increases, modeled as $w_s^m$.
\begin{figure}[t]
	\centering%\captionsetup{width=.5\textwidth}%
	\includegraphics[width=0.5\textwidth]{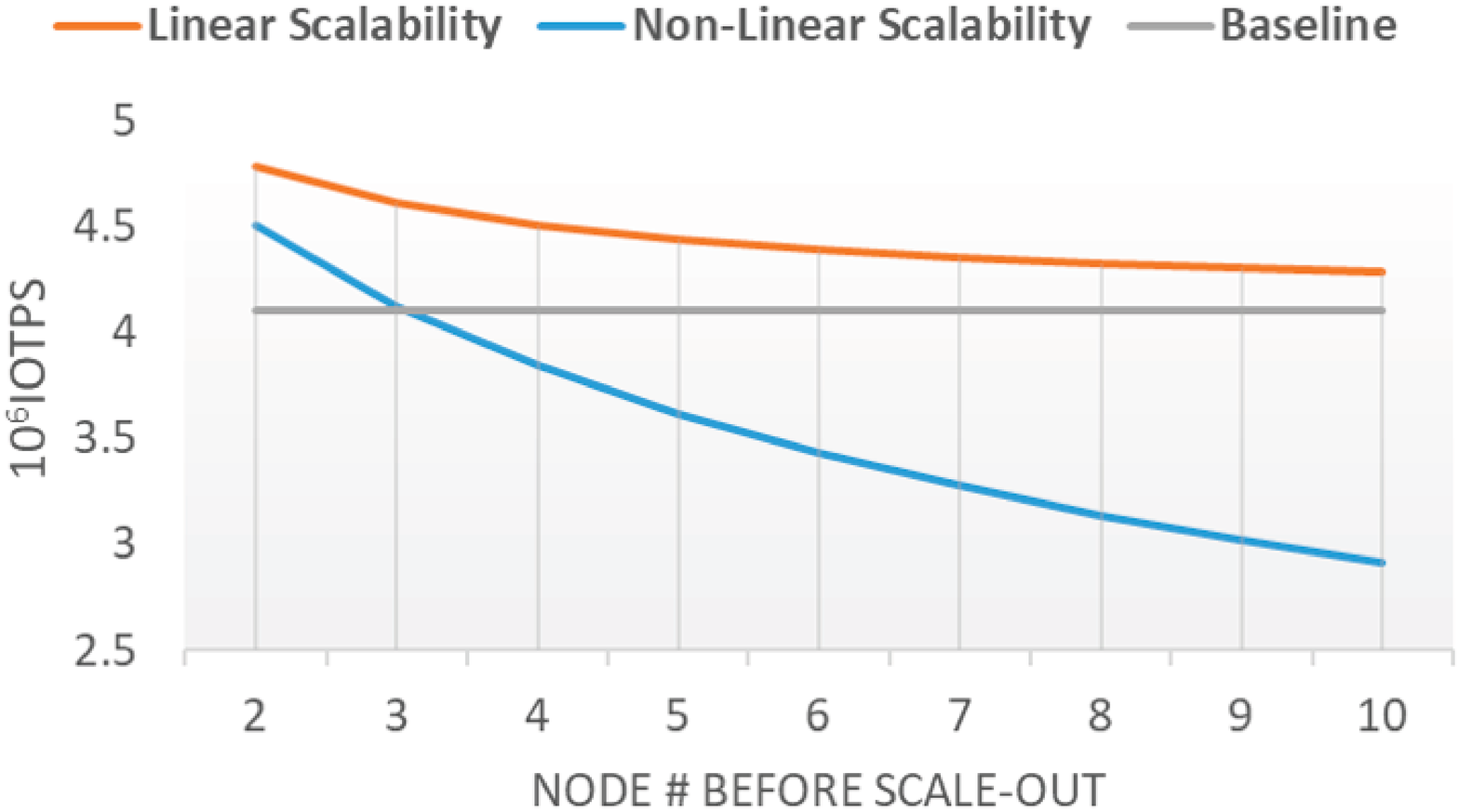}%
	\caption{Performances computed based on Eq.~\eqref{eqn:iotps} for systems with linear and non-linear scalability. The baseline performance in the stable phase is plotted for reference.}\vspace{-12pt}%
	\label{fig:perf} %
\end{figure}

Figure~\ref{fig:perf} presents the results. Assume that a system can achieve $4100$ \emph{KIoTps} before scaling out to one more node. If the system is linear scalable, it can always achieve a performance higher than the baseline. The performance gain from scaling out decreases as the number of system nodes increases. In comparison, systems that are not linearly scalable will have a performance even lower than before scaling out. This is especially true when the system has to migrate data during scale-out process~\cite{dhtlb}. Hence, systems that fail to effectively scale can negatively affect the redesigned metrics, while systems with linear scalability are rewarded in the final metrics.
\begin{figure}[t]
	\centering%\captionsetup{width=.5\textwidth}%
	\includegraphics[width=0.5\textwidth]{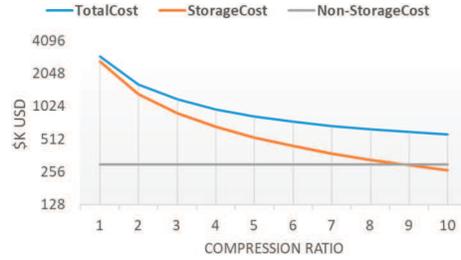}%
	\caption{The storage cost and the total cost of the system, under varied compression ratios.}\vspace{-12pt}%
	\label{fig:syscost} %
\end{figure}
\begin{figure}[t]
	\centering%\captionsetup{width=.5\textwidth}%
	\includegraphics[width=0.5\textwidth]{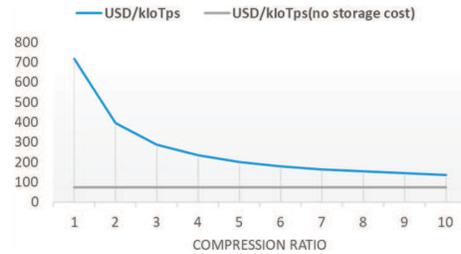}%
	\caption{Price/performance metric, \emph{USD/kIoTps}, under varied compression ratios.}\vspace{-12pt}%
	\label{fig:ppmetric} %
\end{figure}
\subsection{Price/Performance Metric Evaluation}

We evaluate the new prices/performance metric along with the added data compression test. We observe how it can change with different compression ratios. Assuming a system with $4100$ \emph{KIoTps} performance. Using the currently top result of TPCx-IoT as the baseline, we consider a system costing about $\$300K$, with each $16$-byte record making a storage cost of $\$2.039E-08$. Following the new price/performance definition of IoTDataBench, we compute the storage cost and the total cost of  a system running at $4100$ \emph{KIoTps}, under varied compression ratios. Figure~\ref{fig:syscost} plots the result, with the corresponding price/performance results presented in Figure~\ref{fig:ppmetric}.

From Figure~\ref{fig:syscost}, we can see that the storage cost dominates the total cost when the compression ratio is small. On when the compression ratio reaches $10$ can the storage cost be smaller than the non-storage component cost. IoTDataBench assumes one-year preservation of data when designing the price/performance metric. If the data preservation time is longer, the storage cost can be intolerable. This is exactly the reason why users of IoT databases are highly concerned with data compression. The corresponding effects of compression ratio are clearly demonstrated by the price/ performance metric, as shown in Figure~\ref{fig:ppmetric}.\vspace{-6pt}

%\subsection{Data Generation}
%\subsection{Complete Test}

\section{Discussion}

System scalability has key impacts on the performance metric of the system under test. According to \href{https://db-engines.com/en/ranking/time+series+dbms}{DB-Engines}, the most popular time-series databases include InfluxDB~\cite{influxdb}, OpenTSDB~\cite{opentsdb}, KairosDB~\cite{kairosdb}, TimescaleDB~\cite{timescaledb}, and Druid~\cite{druid}. While the details of InfluxDB's distributed version are unknown for its closed source, OpenTSDB, KairosDB, TimescaleDB and Druid are exploiting existent distributed systems for scalability, i.e., HBase~\cite{hbase}, Cassandra~\cite{cassandra}, distributed PostgreSQL database~\cite{pgsql} and HDFS~\cite{hdfs}. These distributed systems are designed for file, key-value, or relational data. IoT time-series data have features and workload characteristics different from these types of data. In fact, the data distribution design of existent scalable systems cannot meet the workload characteristics of time-series data~\cite{dhtlb}. We believe a new data distribution architecture for IoT time-series data is demanded.

As for data compression, existent time-series databases have employed both lossless compression~\cite{gorilla,rle,sprintz,delta,fpzip} and lossy compression~\cite{lfzip,scientificRequirements,bioCompression}. They have also commonly adopted the data retention policy to control storage consumption by discarding data passing a given time~\cite{influxdb}. But discarding historical data is causing a loss~\cite{peregreen}. For example, historical data are crucial for long-term observations and enabling new scientific knowledge creation in the future~\cite{historicData}. Considering the cost of storage, we believe it is desirable to have time-series database that can store as much data as possible within a given storage space.\vspace{-6pt}

\section{Conclusions}

IoT data management is becoming prevalent due to the increase of IoT devices and sensors. As an industrial standard for benchmarking IoT databases, TPCx-IoT is gaining attention gradually. In this paper, we present our practicing experience of TPCx-IoT in achieving a record-breaking result, as well as lessons learned when applying TPCx-IoT to our real-world use case. Driven by users' needs to evaluate data compression and system scalability, we extend TPCx-IoT and propose IoTDataBench, which updates four aspects of TPCx-IoT, i.e., data generation, workloads, metrics and test procedures. Preliminary implementation and evaluation results of IoTDataBench are provided. Based on our design process of IoTDataBench, we summarize two demanding aspects to be addressed by IoT time-series database designers, i.e., how to distribute IoT time-series data that suits its workload characteristics, and how to store as much data as possible within a given storage space.

\section*{Acknowledgements}\vspace{-12pt}

This work was supported by NSFC Grant (No. 62021002).\vspace{-6pt}


%
\bibliographystyle{splncs04}
%{\footnotesize
\bibliography{ref}
%} %
%
\end{document}